\documentclass[12pt, letterpaper]{article}

\usepackage{amsmath,amssymb,amsbsy,amsfonts,amsthm}
\usepackage{alltt}
\usepackage{physics,graphicx,booktabs}
\usepackage{setspace}
\usepackage{caption}
\usepackage{natbib}
\usepackage[colorlinks, citecolor=blue,linkcolor=blue]{hyperref}
\theoremstyle{plain}% Theorem-like structures provided by amsthm.sty

\usepackage[margin=1in]{geometry}

\usepackage{color}
\newcommand{\blue}[1]{{\textcolor{black}{#1}}}

\renewcommand{\hat}{\widehat}
\renewcommand{\bar}{\overline}

\definecolor{fgcolor}{rgb}{0, 0, 0}

\usepackage{framed}
\makeatletter
{\par\unskip\endMakeFramed%
\at@end@of@kframe}
\makeatother

\definecolor{shadecolor}{rgb}{.97, .97, .97}
\definecolor{messagecolor}{rgb}{0, 0, 0}
\definecolor{warningcolor}{rgb}{1, 0, 1}
\definecolor{errorcolor}{rgb}{1, 0, 0}
\theoremstyle{definition}

\theoremstyle{remark}

\begin{document}

%\articletype{ARTICLE TEMPLATE}% Specify the article type or omit as appropriate

\title{Time Fused Coefficient SIR Model with Application to
		COVID-19 Epidemic in the United States}

\author{
Hou-Cheng Yang~~~%\textsuperscript{a}
%\thanks{H.-C. Yang and Y. Xue contributed equally.},
Yishu Xue~~~%\textsuperscript{b}, 
Yuqing Pan~~~%\textsuperscript{c},
Qingyang Liu~~~%\textsuperscript{b}
Guanyu Hu}%\textsuperscript{d}\thanks{CONTACT Guanyu Hu. Email:
%gh7mr@missouri.edu}}
%\affil{\textsuperscript{a}Departement of Statistics, Florida
%State University, Tallahassee, FL; ORCID 0000-0002-8679-4280\\
%\textsuperscript{b}Department of Statistics, University of
%Connecticut, Storrs, CT; ORCID 0000-0002-9660-6087\\
%\textsuperscript{c}Microsoft Corporation, Redmond, WA;\\
%\textsuperscript{d}Department of Statistics, University of
%Missouri Columbia, Columbia, MO; ORCID 0000-0003-1410-1665
%}

\maketitle

\begin{abstract}
In this paper, we propose a Susceptible-Infected-Removal (SIR) model with time
fused coefficients. In particular, our proposed model discovers the underlying
time homogeneity pattern for the SIR model's transmission rate and removal rate
via Bayesian shrinkage priors.
\blue{MCMC sampling for the 
proposed method
is facilitated by the \textbf{nimble} package in \textsf{R}.}
Extensive simulation studies are carried out to examine the
empirical performance of the proposed methods. We further apply the proposed
methodology to analyze different levels of COVID-19 data in the United States.
\end{abstract}

\section{Introduction}

The severe acute respiratory syndrome coronavirus~2 (SARS-CoV-2) was first
identified in December 2019, and then rapidly spread across the world, 
causing the current global pandemic of coronavirus disease 2019 (COVID-19). 
As of July~23, the novel coronavirus has spread to~216 countries and 
territories, with a total of more than~14 million confirmed infections and 
600,000 fatal cases worldwide \citep{who2020}. Eight months after the initial 
outbreak, large numbers of new cases are still reported from many major 
countries, resulting in not only public health crises, but also severe 
economic and political ramifications. As the pandemic rages on with no end in 
sight, it is of urgent necessity for epidemiologists to quantify and 
interpret the trajectories of the COVID-19 pandemic, so as to help formulate 
more effective public policies. 

The Susceptible-Infectious-Recovered \citep[SIR;][]{kermack1927contribution}
model and its variants, such as Susceptible-Infected-Removed-Susceptible 
\citep[SIRS;][]{kermack1932contributions,kermack1933contributions}
and Susceptible-Exposed-Infected-Removal 
\citep[SEIR;][]{hethcote2000mathematics}
models are commonly used to describe the dynamics of an infectious disease in a
certain region. In the basic SIR model, a population is segregated into three
time-dependent compartments including Susceptible ($S(t)$), Infectious
($I(t)$),
and Recovered/removed ($R(t)$). One who does not have the disease at time~$t$,
but may be infected due to contact with an infected person belongs to the
susceptible compartment. The infected compartment is made up of those who have
a
disease at time~$t$, and can potentially get a susceptible individual infected
by contact. The recovered compartment include those who are either recovered or
dead from the disease, and are no longer contagious, i.e., removed from the
infectious compartment, at time~$t$. Removal can be due to several possible
reasons, including death, recovery with immunity against reinfection, and
quarantine and isolation from the rest of the population. A recovered/removed
individual will not be back into the susceptible compartment anymore. Such
model assumption match well with the COVID-19 outbreak, and therefore we adopt
the SIR model as our basic model in this paper.

From the statistical perspective, the key study objective is the inference of
transmission and recovery rates from the model. Regarding time-invariant SIR
and
SEIR models, there have been timely applications to early epidemic data right
after the \blue{breakout of COVID-19}
\citep{read2020novel,tang2020estimation,wu2020nowcasting}. In order to
differentiate evolutional patterns of COVID-19 among different
regions, Hu and Geng \cite{hu2020heterogeneity} developed a Bayesian
heterogeneity learning methodology for SIRS.  As the epidemic continued to
spread rampantly, statisticians proposed time-dependent models based on SIR to
elucidate the temporal dynamics of this disease
\citep{chen2020time,jo2020analysis,sun2020tracking}. Estimated by various
assumptions on temporal smoothness, the transmission and recovery rates of
these
models constantly alter over time, \blue{which limits their ability to
effectively detect abrupt changes.}
In contrast, we consider the scenario in which the transmission and
recovery rates are constant within locally stationary periods segmented by a
collection of change points, which is aligned with the fact that different
stages of epidemic progression are naturally partitioned. This motivates us to
estimate a piecewise constant model.

The fused lasso \cite{tibshirani2005sparsity}, with
$L_1$ sparsity-inducing penalty
imposed on all successive differences, is one of the most popular methods for
time fusion and change point detection. Motivated by the frequentist $L_1$
fusion penalty, Kyung et al. \cite{kyung2010penalized} proposed its Bayesian
counterpart, namely Bayesian fused lasso, \blue{which imposed independent
Laplace priors \citep[][]{park2008bayesian} on the differences}. To
solve the posterior inconsistency problem of Bayesian fused lasso, Song and
Cheng \cite{song2019bayesian} used heavier tailed student-$t$ priors for
Bayesian fusion estimation. In addition to the Laplace and student-$t$ priors,
other
Bayesian shrinkage priors with different statistical properties, such as
spike-and-slab \citep{george1993variable} and  horseshoe priors
\citep{carvalho2010horseshoe}, can also be adopted to induce time fusion.

The contributions of this paper are in three-fold. First, we apply three
different types of shrinkage priors to capture the time homogeneity patterns of
infectious and removal rates under the SIR framework. Second, it is noticed
that
our proposed method can be easily implemented by the \textbf{nimble} package
\cite{de2017programming} in
\textsf{R}. A straightforward tutorial on using \textbf{nimble} to obtain
shrinkage priors under the SIR framework is provided \blue{in the
supplemental material}. Finally, several
interesting findings are discovered through analysis of
COVID-19 data, including
including national level, state level, and county level. 

The remainder of this paper is organized as follows. In
Section~\ref{sec:motivate_data}, the COVID-19 data of selected state and county
are introduced. We \blue{briefly review the} SIR model, and then present our
model framework in Section~\ref{sec:method}.
Simulation studies are conducted in
Section~\ref{sec:simulation}. Applications of the proposed methods to COVID-19
data are presented in Section~\ref{sec:realdata}. Section~\ref{sec:conclusion}
concludes the paper with a discussion.

\section{Motivating Data}\label{sec:motivate_data}

The COVID-19 data is obtained from the~\textsf{R} package~\textbf{COVID19}
\citep{Guidotti2020}. We consider the observations recorded from 2020-05-14 to
2020-07-23, a 71-day long period.
US nationwide aggregated data, as well as data for five states: New York (NY),
California (CA), Florida (FL), South Dakota (SD), and Wyoming (WY) are our
focus
in this study. We also consider the county-level data including: Los Angeles,
Miami-Dade and New York City. The data is reported daily, with variables
including the population size, the number of confirmed cases, the number of
recoveries, and the number of deaths, etc.

\begin{figure}[tbp]
	\centering
	\includegraphics[width=\textwidth]{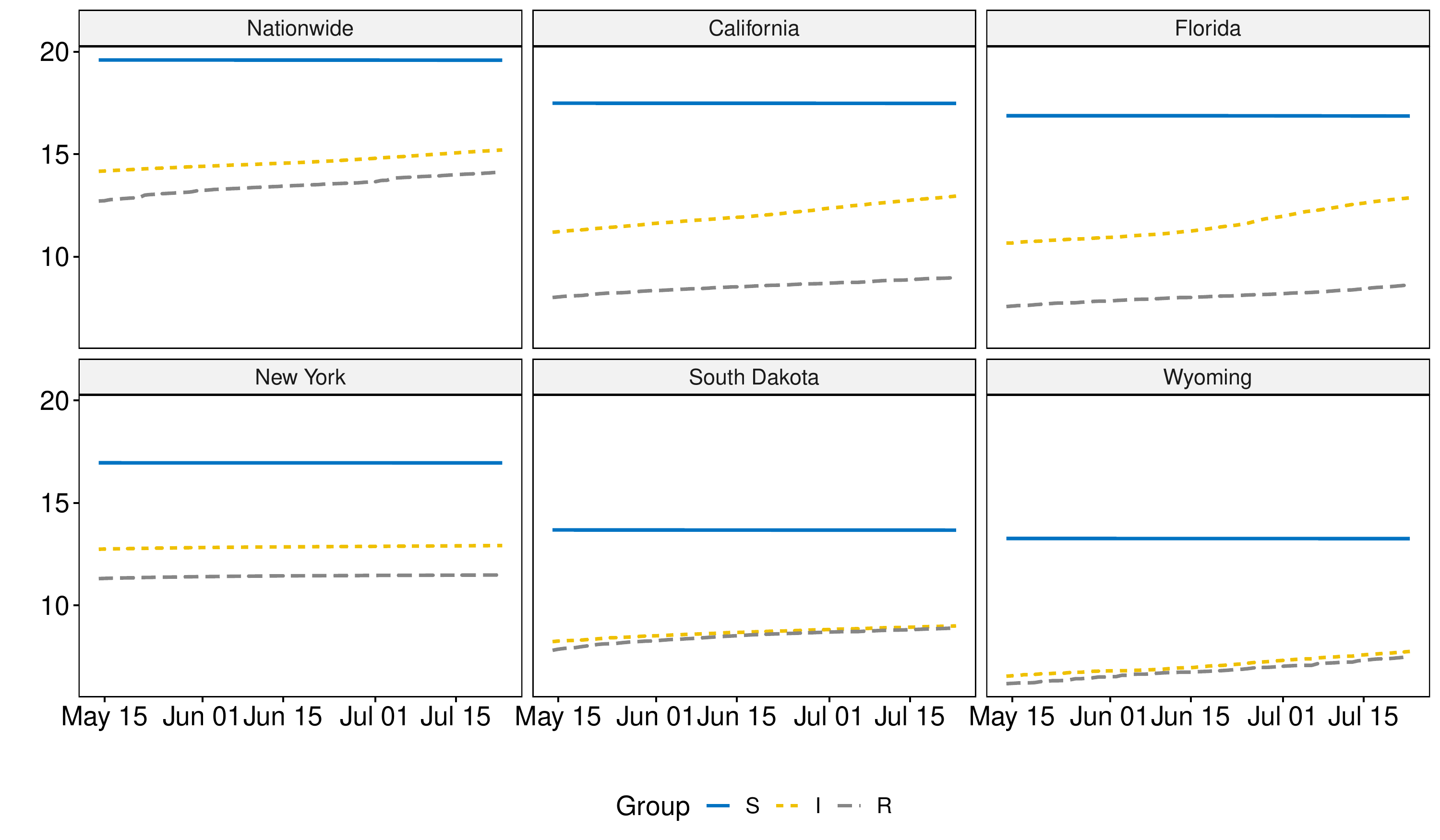}
	\caption{Visualizations for $S(t)$,
		$I(t)$ and $R(t)$ in US nationwide and
		five individual states on natural log scale.}
	\label{fig:state_data}
\end{figure}

\begin{figure}[tbp]
	\centering
	\includegraphics[width=\textwidth]{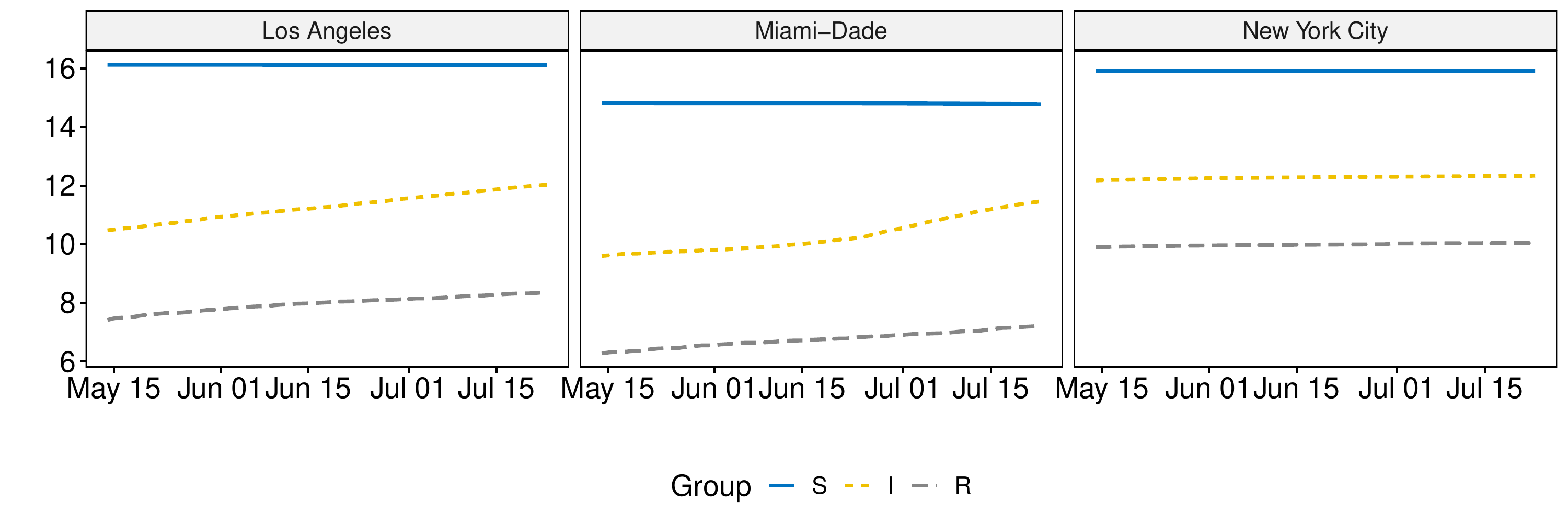}
	\caption{Visualizations for $S(t)$,
		$I(t)$ and $R(t)$ in the three
		selected counties on natural log scale.}
	\label{fig:county_data}
\end{figure}

Note that the removal group for county-level data only contain deaths and there
is no information available for recoveries. Similar to in Sun et al.
\cite{sun2020tracking}, a three-point moving average filter is applied to the
infectious group~$I(t)$ and removal group~$R(t)$ to \blue{reduce} noise. Due to
the
large size of the susceptible group, the group sizes are visualized on a
natural
log scale in Figure~\ref{fig:state_data} and Figure~\ref{fig:county_data}. The
infectious and removal numbers are much smaller in SD and WY when compared to
other states. As during the studied period, NY is still under lock-down, both
the infectious and removal groups experienced slow increases. For CA and FL,
however, potentially due to re-open in early May, their infectious and removal
groups saw rapid increases. The three counties selected are the metropolitan
areas in CA, FL and NY, and the trends observed are similar to those in their
respective states.

\section{Method}\label{sec:method}
\subsection{The SIR and vSIR Models}

In the SIR model, we consider a fixed total population of size~$N$. By
``fixed'', we assume that the population size does not vary over time. The
effect of natural death or birth are not considered here, as the outstanding
period of an infectious disease is much shorter than human average lifetime.
Denote, at time~$t$ ($t \geq 1$), the counts of susceptible, infectious, and
recovered/removed persons within a given region as~$S(t)$, $I(t)$ and~$R(t)$,
respectively, and the relationship $N = S(t) + I(t) + R(t)$ always holds.
%\blue{Und}
%The focus of this study under SIR framework is how the size of each group
%change
% over time.

Two parameters in the SIR model are time-invariant: the transmission
rate~$\beta$, and the recovering rate~$\gamma$. The transmission rate~$\beta$
controls how much the disease can be transmitted through exposure. It is
jointly
determined by the chance of contact and the probability of disease
transmission.
The recovering rate~$\gamma$ stands for the rate at which infected individuals
recover or die. Time-varying property of these two parameters is ignored in
traditional SIR modeling, which is a rather strong simplifying assumption that
hurdles the model's prediction power for disease trend. Therefore, we adopt the
time-varying SIR \citep[vSIR;][]{sun2020tracking} framework, where both~$\beta$
and~$\gamma$ are functions of time~$t$.

The vSIR model can be viewed as both a deterministic model and a stochastic
model. The deterministic vSIR model allows us to describe the number of people
in each compartment with the ordinary differential equations (ODEs). A
generalized version of the deterministic vSIR model with infectious
rate~$\beta(t)$ and removal rate~$\gamma(t)$ respect to time
\blue{can be described} as follows:
\begin{align}
\nonumber
	&\frac{\dd S(t)}{\dd t} = \frac{-\beta(t)I(t)S(t)}{N},\\
\nonumber                 
	&\frac{\dd I(t)}{\dd t} = \frac{\beta(t)I(t)S(t)}{N}  - \gamma(t)I(t),\\      
	&\frac{\dd R(t)}{\dd t}  = \gamma(t)I(t).                    
\end{align}
While the deterministic vSIR model seems appealing due to its simplicity, the
spread of a disease, however, is naturally stochastic. Disease transmission
between two individuals is random rather than deterministic. The stochastic
formulation of the vSIR model is, therefore, preferred for epidemic modeling
purposes, as it allows for randomness in the disease spreading process.

\subsection{Time Fusion SIR Model}\label{ssec:hiemodel}

Consider the the vSIR-Poisson process framework of \cite{sun2020tracking} with
two time-varying parameters, $\beta(t)$ and~$\gamma(t)$.
Let~$N$ be the total population,
$M(t)=I(t)+R(t)$ denote the cumulative number of diagnosed cases
and~$\Delta{M(t)}=M(t)-M(t-1)$, $\Delta{R(t)}=R(t)-R(t-1)$ represent the daily
changes of~$M(t)$ and~$R(t)$.
The initial values $\Delta M(1)$ and $\Delta R(1)$ are defaulted, respectively,
to~$M(1)$ and $R(1)$.
Let $t=1,\ldots,T$ denote the time domain.
Hence we have
\begin{align}
\nonumber
\Delta{M(t)}& \sim \text{Poisson}\left(\frac{\beta(t)S(t)I(t)}{N}\right),\\
\Delta{R(t)}& \sim \text{Poisson}\left(\gamma(t)I(t)\right), ~~
t=\blue{2},\ldots, T.
\end{align}
%Then we focus on Bayesian fusion of the successive differences~$\Delta
%\beta(t)$ and~$\Delta \gamma(t)$.
%As the model focuses on time fusion, we name it as hierarchical time
%fusion SIR (tf-SIR).
For most infectious diseases, the infectious and removal rates~$\beta(t)$
and~$\gamma(t)$ do not always change smoothly over time, as they can be
influenced by certain government policies in a notable manner. In other words,
$\beta(t)$ and~$\gamma(t)$ can fluctuate around a fixed value within a specific
time period, and then with the inception of a policy, fluctuate around another
value in a period that follows. Identifying the subpopulation structure of these
two parameters with time fusion patterns in the SIR model will enhance our
understanding of infectious diseases such as COVID-19. In this paper, we assume
that successive differences of the infectious rate
$\Delta\beta(t)=\beta(t)-\beta(t-1)$ and removal rate
$\Delta\gamma(t)=\gamma(t)-\gamma(t-1)$ both have an unknown clustered pattern
with respect to time. Both~$\Delta \beta(1)$ and~$\Delta \gamma(1)$ are
defaulted to~0. For example, with a cluster of 0's in the successive
differences, $\beta(t)$ would remain constant over the corresponding time
period. Towards this end, we use three different shrinkage priors on
both~$\Delta \beta(t)$ and $\Delta \gamma(t)$ to detect such clusters, including
the student-$t$ prior, horseshoe prior and spike-and-slab prior \citep[see,][for
more discussion]{piironen2017sparsity,song2019bayesian}. As our proposed model
focuses on time fusion, we name it hierarchical time fusion SIR (tf-SIR).

The first prior we consider is student-$t$ prior. 
Despite the popularity of the Laplace prior, it has a light tail,
suffers from posterior inconsistency issues
\citep{song2017nearly,song2019bayesian}, and
often leads to smoothly varying estimation results, i.e., it cannot identify
the clustered structure. The student-$t$ prior, with its heavier tail, induces
stronger shrinkage effect, and enjoys a nice posterior consistency property.
%Such a fusion prior will 
%shrink successive differences towards zero and hence induce posterior
%blocking.
%Compared to Laplace fusion prior, student-$t$ fusion prior induces stronger
%shrinkage effect and enjoys a nice posterior consistency property
%\citep{song2019bayesian}. As Laplace prior suffers from
%posterior inconsistency due to
%its light tail, Song and Liang \cite{song2017nearly} suggested using heavy
%tail
%distributions, e.g., student-$t$ prior. In addition,
%the Laplace prior usually leads to
%a smoothly varying estimation, which means it cannot identify the clustered
%structure. 
The student-$t$ shrinkage prior on the successive differences
can be written as:
\begin{align*}
& \blue{\beta(t) - \beta(t-1)} \mid \sigma^2_\beta \sim t_{df_\beta}(l_\beta
\sigma_\beta),~\sigma_\beta^2 \sim \mbox{IG}(a_{\sigma_\beta},
b_{\sigma_\beta}), ~ \beta(1)\mid \sigma_\beta^2 \sim
\mbox{N}(0,\sigma_\beta^2 \lambda_1),\\
& \blue{\gamma(t) - \gamma(t-1)}  \mid \sigma^2_\gamma \sim
t_{df_\gamma}(l_\gamma
\sigma_\gamma),~\sigma_\gamma^2 \sim \mbox{IG}(a_{\sigma_\gamma},
b_{\sigma_\gamma}), ~ \gamma(1)\mid \sigma_\gamma^2 \sim
\mbox{N}(0,\sigma_\gamma^2 \eta_1) ,~t=2,\ldots, T,
\end{align*}
%\begin{align}
%\nonumber
%&\Delta\beta(t)\mid\sigma^2_\beta\sim
%t_{df_\beta}\left(l_\beta\sigma_\beta\right), ~~
%\sigma^2_\beta\sim\text{IG}\left(a_{\sigma_\beta}, b_{\sigma_\beta}\right),
%~~
%\beta
%(1)\mid\sigma^2_\beta\sim\text{N}\left(0,\sigma^2_\beta\lambda_1\right),~~
%t=2,\ldots, T,
%\end{align}
where~$T$ denotes the termination time of
observation, $t_{\omega_1}\left(\omega_2\right)$ denotes the
{student-$t$} distribution with degree of freedom~$\omega_1$ and scale
parameter~$\omega_2$, N() stands for the normal distribution, and IG()
stands for the inverse gamma distribution.
Note that the above {student-$t$} distribution can be
rewritten as an inverse gamma scaled Gaussian mixture, and hence the
tf-SIR model with student-$t$ prior \blue{for the sequential differences
in~$\beta(t)$ and~$\gamma(t)$}
 can be alternatively formulated as:

\begin{align}\label{eq:tshrink}
% \nonumber
% &\Delta{M(t)}\sim \text{Poisson}\left(\frac{\beta(t)S(t)I(t)}{N}\right),\\
% \nonumber
% &\Delta{R(t)}\sim \text{Poisson}\left(\gamma(t)I(t)\right),\\
\nonumber
&\Delta \beta(t) \mid\sigma^2_\beta,
\lambda_t\sim\text{N}(0,\lambda_t\sigma^2_\beta), \quad
\lambda_t\sim\text{IG}\left(a, b\right),\\
\nonumber
&\Delta \gamma(t) \mid\sigma^2_\gamma,
\eta_t\sim\text{N}(0,\eta_t\sigma^2_\gamma), \quad
\eta_t\sim\text{IG}\left(c, d\right),\\
&\sigma^2_\beta\sim\text{IG}\left(a_{\sigma_\beta}, b_{\sigma_\beta}\right),
\quad \sigma^2_\gamma\sim\text{IG}\left(a_{\sigma_\gamma},
b_{\sigma_\gamma}\right), \quad t=2,\ldots, T,
\end{align}
where $a$, $b$ satisfy conditions $df_\beta=2a$ and $l_\beta =
\sqrt{\frac{b}{a}}$. Similarly, $c$, $d$ satisfy conditions
$df_\gamma=2c$ and $l_\gamma = \sqrt{\frac{d}{c}}$.
\blue{Both~$\sigma_\beta^2$ and~$\sigma_\gamma^2$ are fixed,
global parameters that shrink the successive differences towards~0.
Following common practices, we assume inverse gamma
priors for both to impart heavy tails, and keep the probability distribution
further from~0 than the Gamma distribution. Different levels of sparsity can be
achieved by varying the values of~$\sigma_\beta^2$ and~$\sigma_\gamma^2$, with
smaller values inducing stronger shrinkage towards~0.
}

%\begin{align}
%\nonumber
%&\Delta\beta(t)\mid\sigma^2_\beta,
%\lambda_t\sim\text{N}(0,\lambda_t\sigma^2_\beta), \quad
%\lambda_t\sim\text{IG}\left(a_t, b_t\right), \quad
%\sigma^2_\beta\sim\text{IG}\left(a_{\sigma_\beta}, b_{\sigma_\beta}\right), 
%\end{align}
%where \blue{$a_t = df_\beta/2$ and $b_t = l_\beta^2 df_\beta / 2$}.

The second prior is the horseshoe prior
\citep{carvalho2009handling,carvalho2010horseshoe}, which is a continuous
shrinkage prior, and is one of the so called global-local shrinkage prior.
It has exhibited ideal theoretical characteristics, and demonstrated
good empirical performance \citep{datta2013asymptotic,van2014horseshoe}.
%\blue{In our tf-SIR model setting, it is formulated as}
%\begin{align}
%	&\Delta\beta(t)\sim\text{N}(0,c^2\lambda_t^2), \quad
%\lambda_t\sim\text{C}^{+}(0,1),\quad t=2,\ldots, T,
%	\nonumber
%\end{align}
%where $c$ is a global hyperparameter which shrinks all parameters towards
%zero, while the heavy-tailed half-Cauchy prior for the local
%hyperparameter~$\lambda_t$ allows for a subset elements of~$\Delta\beta(t)$ to
%escape from the shrinkage. 
Our tf-SIR model with the horseshoe prior can be expressed as:
\begin{align}
% \nonumber
% &\Delta{M(t)}\sim \text{Poisson}\left(\frac{\beta(t)S(t)I(t)}{N}\right)\\
% \nonumber
% &\Delta{R(t)}\sim \text{Poisson}\left(\gamma(t)I(t)\right)\\
\nonumber
&\beta(t)-\beta(t-1)\mid\sigma^2_\beta,
\lambda_t\sim\text{N}(0,\lambda^2_t\sigma^2_\beta), \quad
\lambda_t\sim\text{C}^{+}\left(0, 1\right),\\
\nonumber
&\gamma(t)-\gamma(t-1)\mid\sigma^2_\gamma,
\eta_t\sim\text{N}(0,\eta^2_t\sigma^2_\gamma), \quad
\eta_t\sim\text{C}^{+}\left(0, 1\right),\\
&\sigma^2_\beta\sim\text{IG}\left(a_{\sigma_\beta}, b_{\sigma_\beta}\right),
\quad \sigma^2_\gamma\sim\text{IG}\left(a_{\sigma_\gamma},
b_{\sigma_\gamma}\right),\quad t=2,\ldots, T,
\end{align}
where $\sigma_\beta^2$ and $\sigma_\gamma^2$ \blue{are same as defined above,}
% are both fixed, global parameters
%shrink all $\Delta \beta_t$ and $\Delta \gamma_t$ towards~0, 
and
$\lambda_t$ and $\eta_t$ are both local parameters following the half-Cauchy
distribution $\mbox{C}^{+}(0,1)$ that allows, respectively,
some $\Delta \beta(t)$ and $\Delta \gamma(t)$ to escape from the shrinkage.
%Note that different levels of sparsity can be achieved by varying the values
%of~$\sigma_\beta^2$ and~$\sigma_\gamma^2$, with smaller values inducing
%stronger shrinkage towards~0.
%Also, the horseshoe prior has exhibited ideal
%theoretical characteristics and good empirical performance

Finally, our third prior of choice is the spike-and-slab prior
\citep{mitchell1988bayesian,george1993variable}. It is a two component discrete
mixture prior. In this paper, we write it as a two-component mixture of
Gaussian
distributions, and the model is expressed as:
\begin{align}\label{eq:spikeslab}
\nonumber
% &\Delta{M(t)}\sim \text{Poisson}\left(\frac{\beta(t)S(t)I(t)}{N}\right),\\
% \nonumber
% &\Delta{R(t)}\sim \text{Poisson}\left(\gamma(t)I(t)\right),\\
\nonumber
&\beta(t)-\beta(t-1)\mid\sigma^2_\beta,  \lambda_t\sim
\lambda_t\text{N}(0,\sigma^2_\beta)+(1-\lambda_t)\text{N}(0,\epsilon^2), \quad
\lambda_t\sim\text{Ber}\left(p\right),\\ %, \quad p:fixed\\
\nonumber
&\gamma(t)-\gamma(t-1)\mid\sigma^2_\gamma,  \eta_t\sim
\eta_t\text{N}(0,\sigma^2_\gamma)+(1-\eta_t)\text{N}(0,\epsilon^2), \quad
\eta_t\sim\text{Ber}\left(\pi\right),\\ %, \quad \pi:fixed\\
&\sigma^2_\beta\sim\text{IG}\left(a_{\sigma_\beta}, b_{\sigma_\beta}\right),
\quad \sigma^2_\gamma\sim\text{IG}\left(a_{\sigma_\gamma},
b_{\sigma_\gamma}\right),\quad t=2,\ldots, T,
\end{align}
where~$\epsilon \ll \sigma_\beta^2$ and $\epsilon \ll \sigma_\gamma^2$,
$\lambda_t$ and~$\eta_t$ are indicators that take values in~$\{0,1\}$,
and~$\mbox{Ber()}$ denotes the Bernoulli distribution. In this paper, we fix
the inclusion probabilities~$p$ and~$\pi$.
%For both~$\sigma_\beta^2$
%and~$\sigma_\gamma^2$, we assume inverse gamma priors to impart heavy tails,
%and keep the probability distribution further from~0 than the Gamma
%distribution.
%$\epsilon\ll c$, and \blue{$\lambda_t$} is an indicator that takes values
%in $\{0, 1\}$. 
In some cases,~$\epsilon$ is set to~0 so that the spike is taken
to a point mass at the origin~$\delta_0$. This distribution can be sensitive to
prior choices of the slab width or prior inclusion probability, and
therefore we choose the normal distribution with a small variance centered
at~0 as the spike.

\section{Simulation}\label{sec:simulation}

\subsection{Simulation Designs}

\begin{figure}[tbp]
	\centering
	\includegraphics[width=\textwidth]{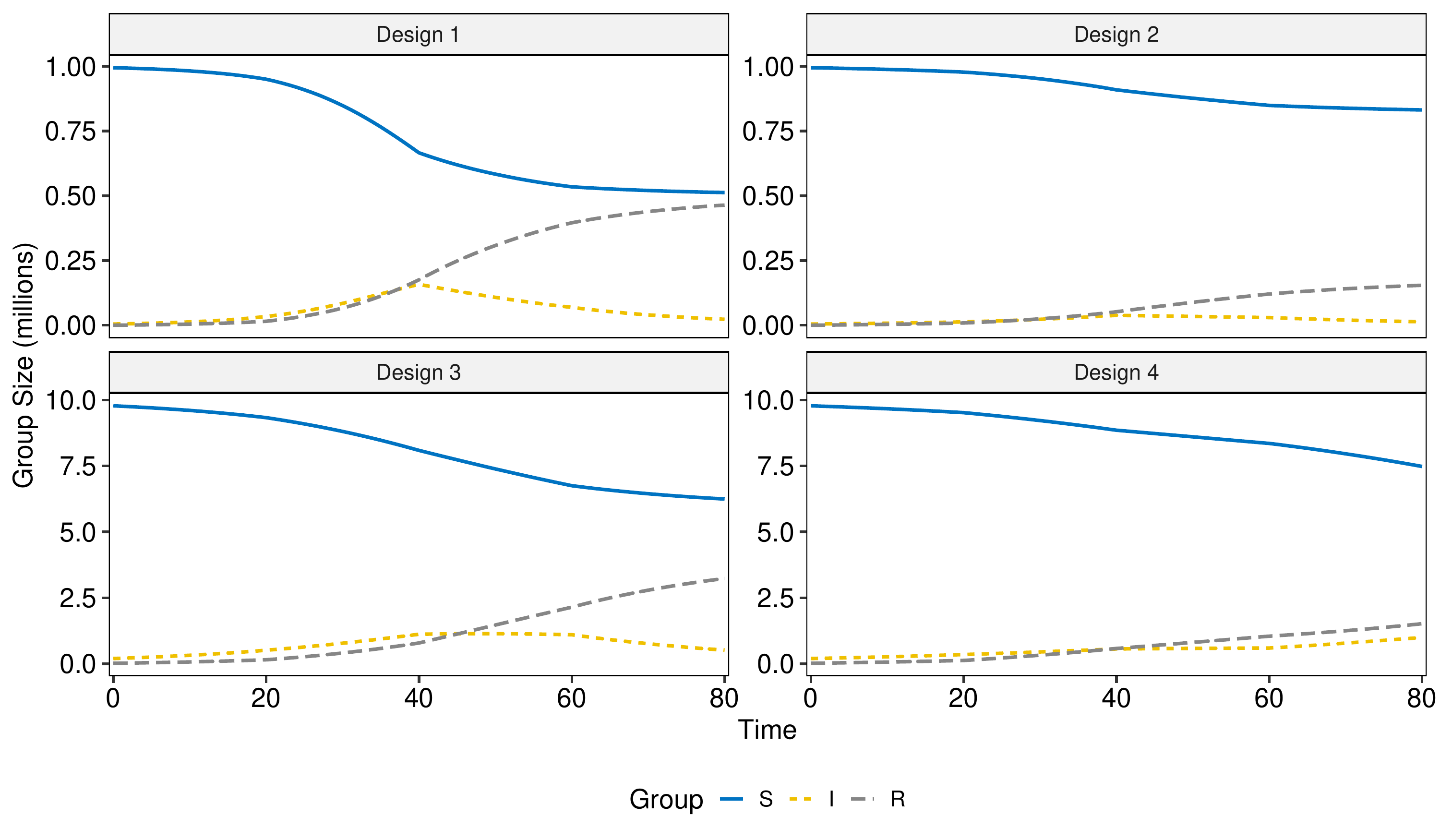}
	\caption{Visualization of example datasets generated under each of the two
	simulation designs.}\label{fig:exampleDatavis}
\end{figure}

We use the \textsf{R} package \textbf{SimInf} \citep{Widgren2019} to generate
data. Four designs over a time span of 80 days are considered. The time domain
is divided into four equally sized pieces each spanning for~20 days, where both
the infectious rate~$\beta(t)$ and the removal rate~$\gamma(t)$ are piecewise
constant within each of them. The four designs have different $\beta(t)$,
$\gamma(t)$, and population size~$N$, and the numerical values are listed in
Table~\ref{tab:parameters}. Four example datasets, one for each design, are
visualized in Figure~\ref{fig:exampleDatavis}. Design~1 corresponds to a fairly
high infectious disease with high removal/recovery rate and design~2
corresponds
to a mildly infectious disease with similar removal/recovery rate. Designs~3
and~4 have larger population sizes, and a disease with smaller numerical values
for~$\beta(t)$ can infect a large portion of the population, such as
demonstrated for Design~3. Design~4, with small~$\beta(t)$ and~$\gamma(t)$,
exhibits slow overall development. A total of~100 replicates are performed for
each design. In each replicate, the length of the MCMC chain is set to~50,000.
\blue{As the numerical values for~$\beta(t)$
and~$\gamma(t)$ in all four designs are not large, it is essential that we get
independent posterior samples. To ensure minimal correlation between draws, we
set the thinning interval to~10 and set the burn-in to 3,000, which leaves us
2,000 samples to perform inference.}

\begin{table}[tbp]
	\centering
	\caption{Parameters used in data generation under the four simulation
	settings.}\label{tab:parameters}
	\begin{tabular}{lccc}
	\toprule
		Design & $\beta(t)$ on pieces 1, 2, 3, 4 &
					$\gamma(t)$ on pieces 1, 2, 3, 4 & Population Size\\
		\midrule
		Design 1 & (0.15, 0.20, 0.10, 0.05) & (0.05, 0.09, 0.10, 0.08) & $10^6$\\
		Design 2 & (0.10, 0.15, 0.10, 0.05) & (0.05, 0.09, 0.10, 0.08) & $10^6$\\
		Design 3 & (0.07, 0.09, 0.08, 0.05) & (0.02, 0.04, 0.06, 0.07) & $10^7$\\
		Design 4 & (0.05, 0.08, 0.05, 0.07) & (0.02, 0.05, 0.04, 0.03) & $10^7$\\
		\bottomrule
	\end{tabular}
\end{table}

\subsection{Performance Measures}

\begin{figure}[tbp]
\centering
\includegraphics[width=\textwidth]{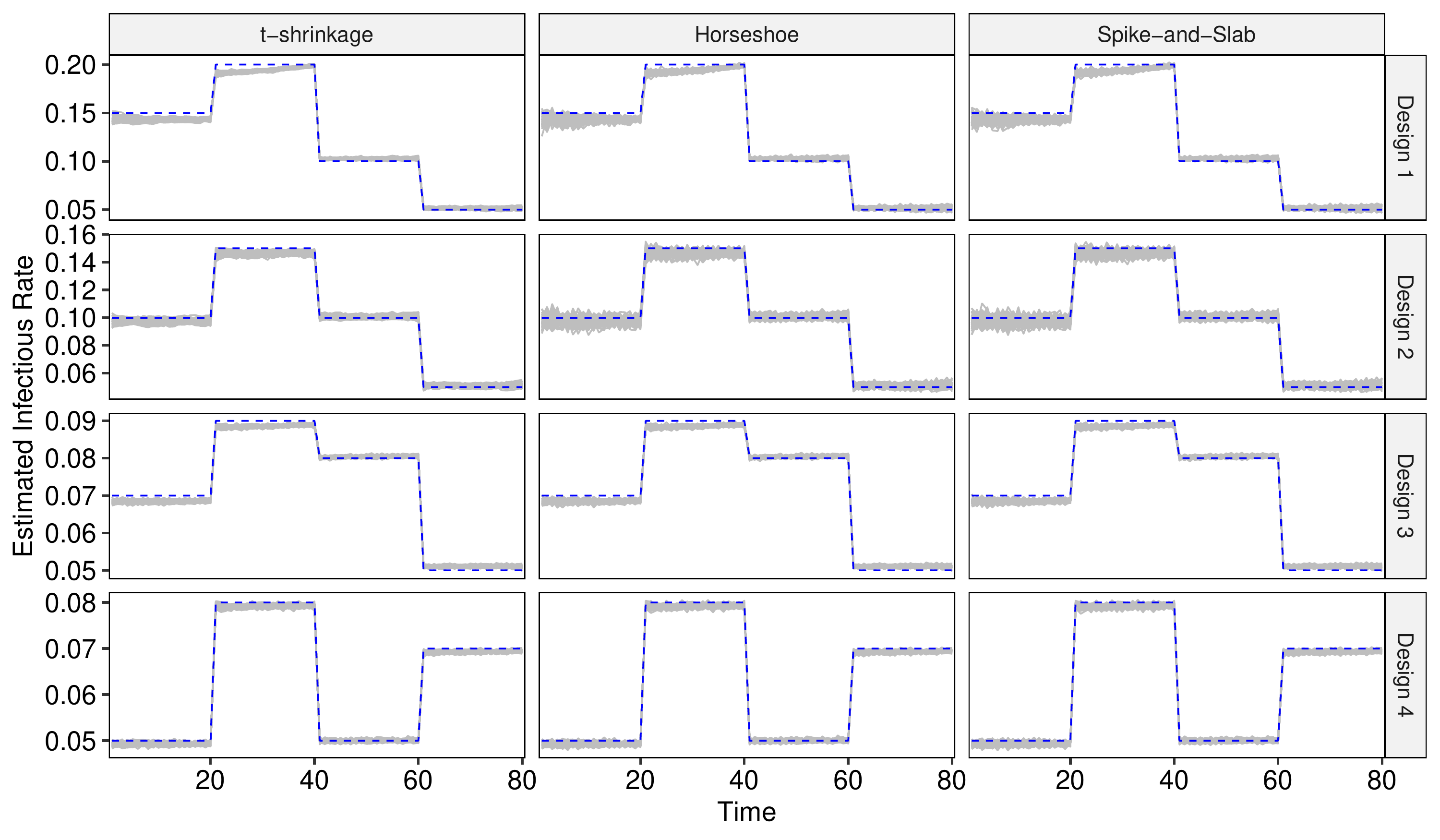}
\caption{Plot of estimated~$\beta(t)$ in~100 replicates for each combination of
design and prior. True values are overlaid in (blue) dashed lines.}
\label{fig:simu_result_beta}
\end{figure}

\begin{figure}[tbp]
\centering
\includegraphics[width=\textwidth]{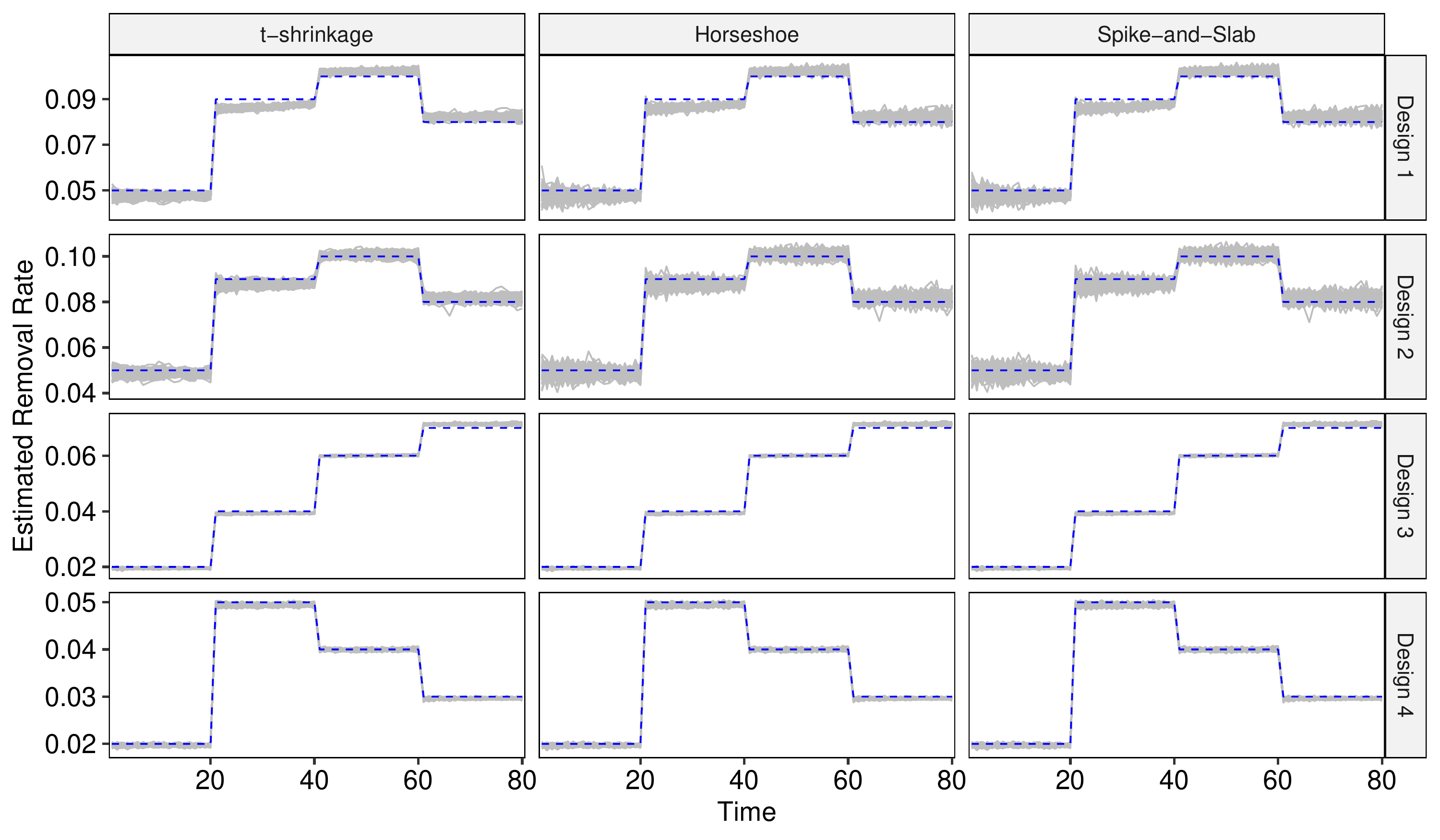}
\caption{Plot of estimated~$\gamma(t)$ in~100 replicates for each combination
of
design and prior. True values are overlaid in (blue) dashed lines.}
\label{fig:simu_result_gamma}
\end{figure}

The parameter estimates for~$\beta(t)$ and~$\gamma(t)$ for the~100 replicates
are visualized respectively in Figures~\ref{fig:simu_result_beta}
and~\ref{fig:simu_result_gamma} as grey lines, and
the true underlying values are also plotted in (blue) dashed lines.
The true underlying values are also plotted in (blue) dashed lines in
Figures~\ref{fig:simu_result_beta} and~\ref{fig:simu_result_gamma}. The first
observation is that, under all four designs, all three models yield quite
accurate parameter estimation performance, as the grey band formed by~100
parameters lie close to or around the (blue) dashed line in both plots.
Secondly, as the population size in Design~3 and Design~4 is~10 times that in
Design~1 and~2, their corresponding grey bands are, overall, tighter. Thirdly,
in both figures, the grey bands corresponding to the $t$-shrinkage prior is
narrower than those for horseshoe and spike-and-slab, indicating overall
relatively stable estimation performance.

The estimation performance, \blue{in addition to visually, is also measured
numerically.} For~$\beta(t)$, \blue{we apply the following three metrics:}
\begin{eqnarray}
	\mbox{MAB}\blue{_\beta}(t) &=& \frac{1}{100}\sum_{\ell=1}^{100}
					\abs{\hat{\beta}_\ell(t) - \beta(t)}, \\
	\mbox{MSE}\blue{_\beta}(t) &=& \frac{1}{100}\sum_{\ell=1}^{100} \left(
	\hat{\beta}_\ell(t) - \beta(t)
	\right)^2,\\
	\mbox{SD}\blue{_\beta}(t) &=& \frac{1}{99}\sum_{\ell=1}^{100}\left(
	\hat{\beta}_\ell(t) - \bar{\hat{\beta}}(t)
	\right)^2,
\end{eqnarray}
where $\hat{\beta}_\ell(t)$ is the posterior estimate of~$\beta$ at time~$t$ in
the~$\ell$th replicate for~$\ell=1,\ldots,100$ and~$t=1,\ldots, 80$, and
$\bar{\hat{\beta}}(t) = \frac{1}{100} \sum_{\ell=1}^{100}\hat{\beta}_\ell(t)$.
\blue{The metrics for $\gamma$ are defined in a similar manner, and therefore
we omit the details.}

%
%This is also reflected in the plots
%for performance measures in Figures~\ref{fig:performance_beta}
%and~\ref{fig:performance_gamma} under the panels for~SD.
%

\begin{figure}[tbp]
	\centering
	\includegraphics[width=\textwidth]{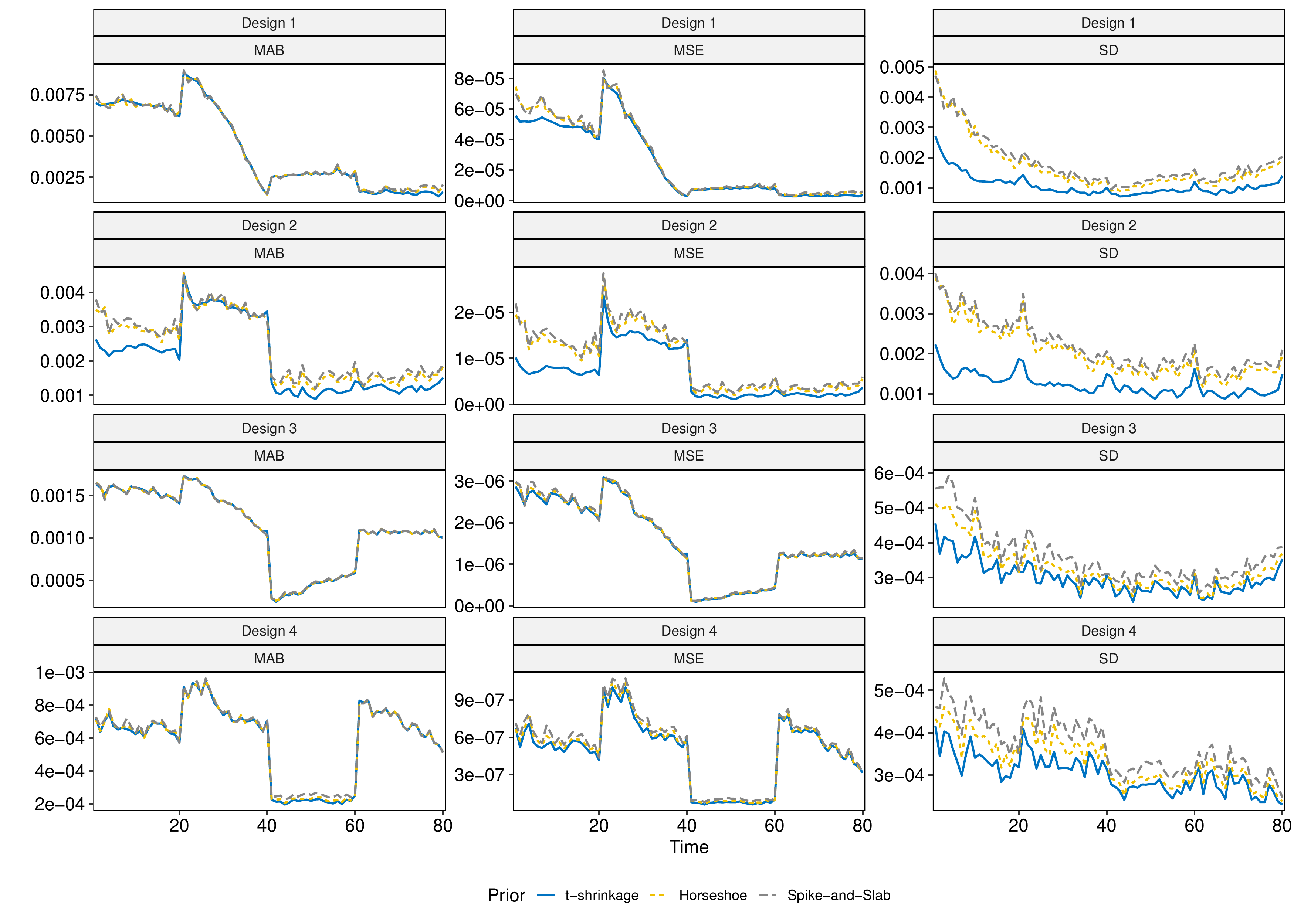}
	\caption{Plot of MAB, MSE and SD of parameter estimate for~$\beta(t)$ under
	different designs.}
	\label{fig:performance_beta}
\end{figure}

\begin{figure}[tbp]
	\centering
	\includegraphics[width=\textwidth]{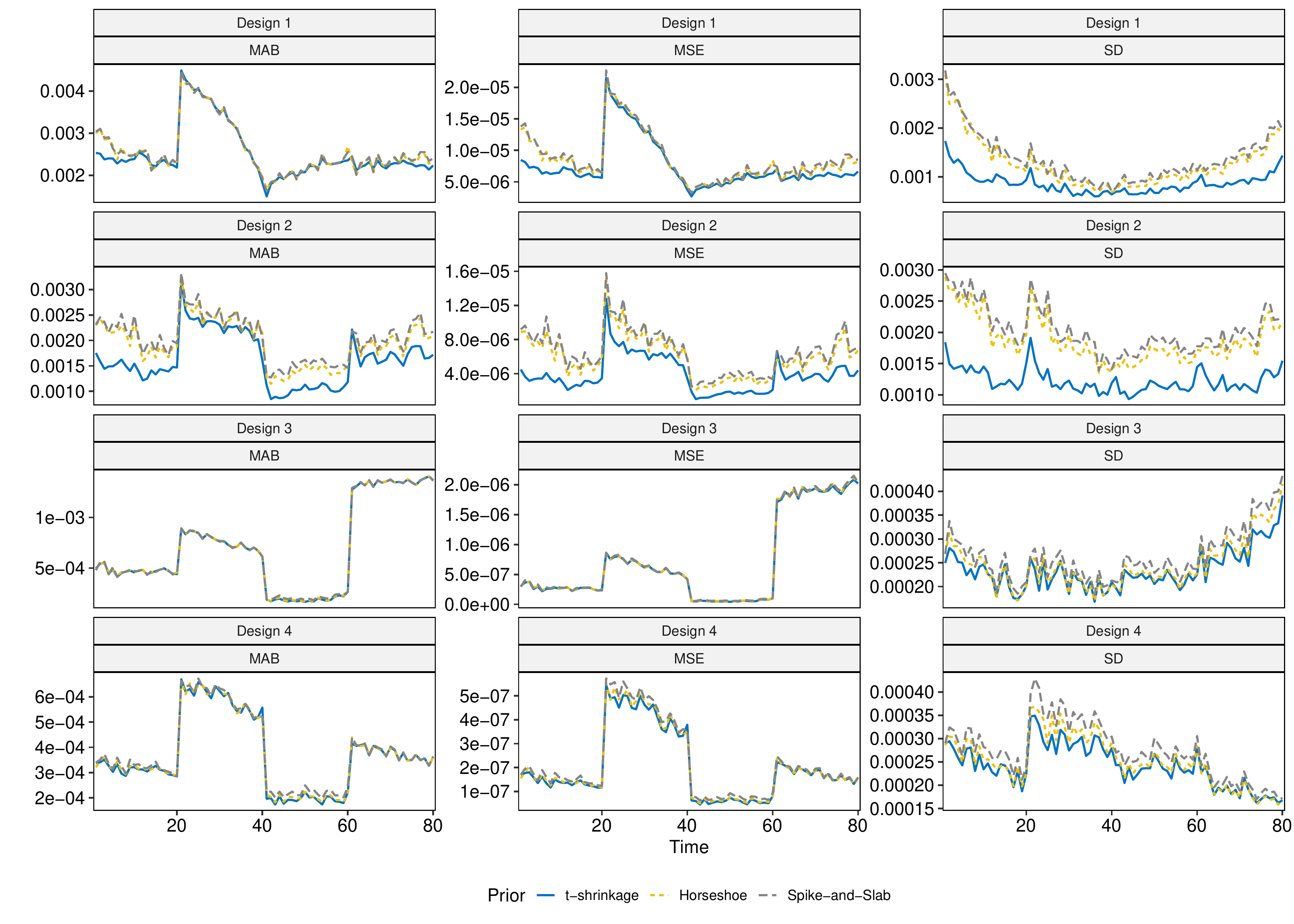}
	\caption{Plot of MAB, MSE and SD of parameter estimate for~$\gamma(t)$ under
	different designs.}
	\label{fig:performance_gamma}
\end{figure}

The three models are compared in terms of the \blue{three}
performance metrics in
Figures~\ref{fig:performance_beta} and~\ref{fig:performance_gamma}. One
interesting observation is that the MAB and MSE tend to be large near when
$t\in\{20,40,60\}$, which corresponds    to when changes in parameters occur.
They then stabilize as the disease continues to develop. For relatively larger
values of the true parameter, the MAB and MSE are larger than for small values
of true parameters. As can be observed from the third column in both
Figures~\ref{fig:performance_beta} and~\ref{fig:performance_gamma}, the
horseshoe and spike-and-slab priors perform similarly in terms of MAB, MSE and
SD. When the sample size is $10^6$, the $t$-shrinkage prior yields parameter
estimates that are overall more stable and have smaller SD than the other two,
\blue{which is consistent with the third observation for the grey bands.}
This difference, however, decreases with increase in sample size. 

\section{Real Data Analysis}\label{sec:realdata}

\begin{figure}[tbp]
	\centering
	\includegraphics[width=\textwidth]{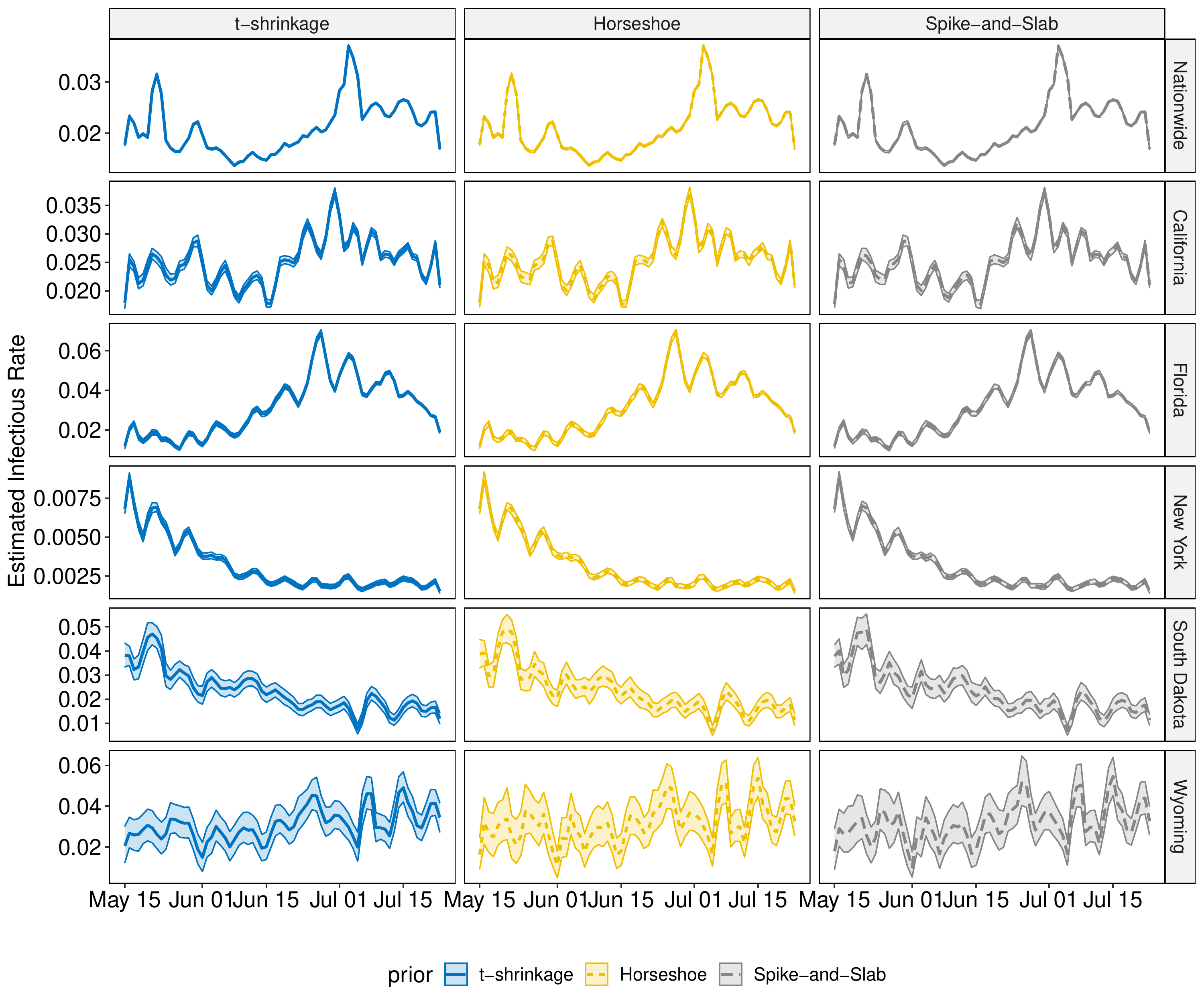}
	\caption{Plot for the estimated infectious rate~$\beta(t)$
	\blue{with 95\% HPD intervals}
	for US nationwide and five individual states over the studied
	time frame.}
	\label{fig:real_beta}
\end{figure}

\begin{figure}[tbp]
	\centering
	\includegraphics[width=\textwidth]{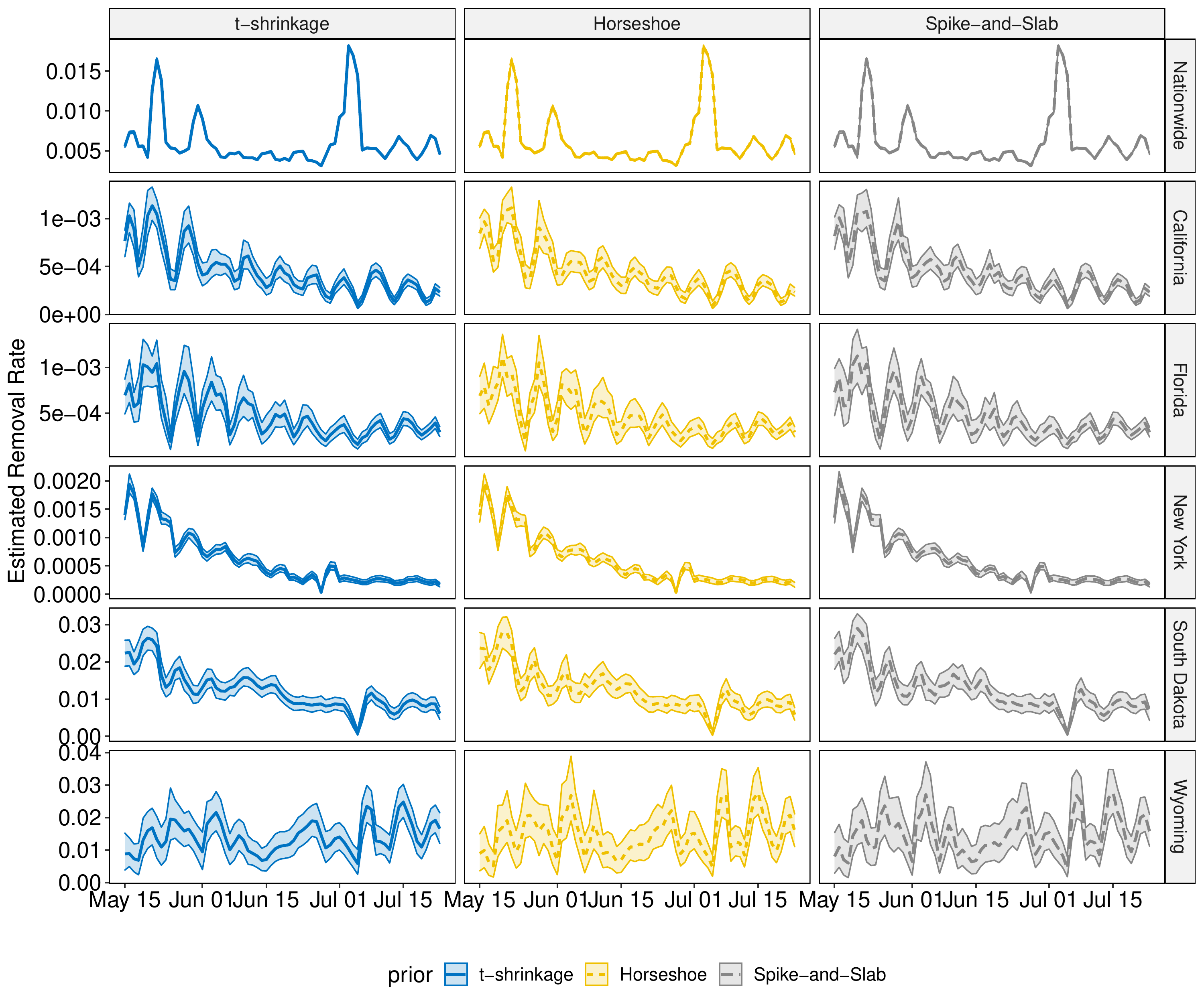}
	\caption{Plot for the estimated removal rate~$\gamma(t)$
	\blue{with 95\% HPD intervals}
	for US nationwide and five individual states over the studied
	time frame.}
	\label{fig:real_gamma}
\end{figure}

\begin{figure}[tbp]
	\centering
	\includegraphics[width=\textwidth]{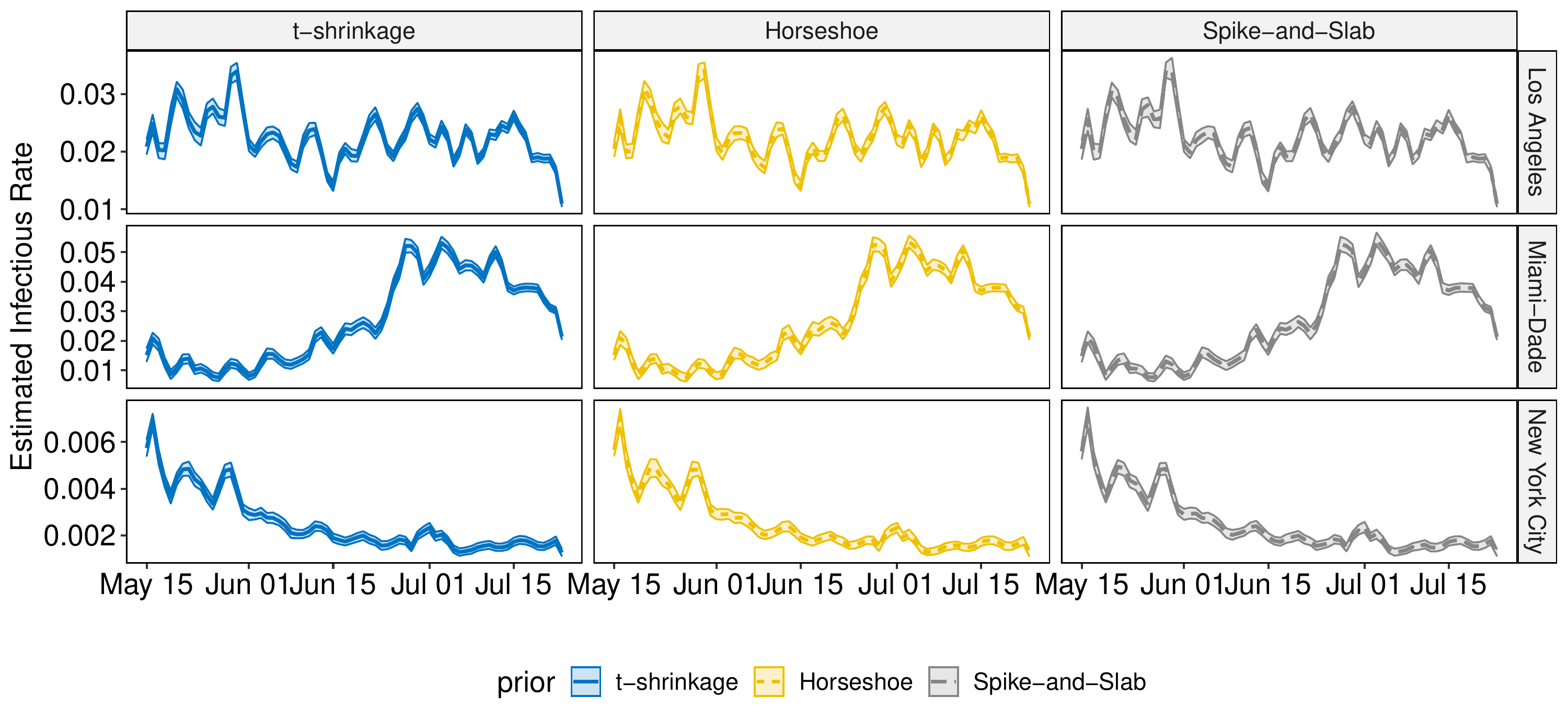}
	\caption{Plot for the estimated infectious rate~$\beta(t)$
	\blue{with 95\% HPD intervals}
	for the three selected counties over the studied
	time frame.}
	\label{fig:county_beta}
\end{figure}

\begin{figure}[tbp]
	\centering
	\includegraphics[width=\textwidth]{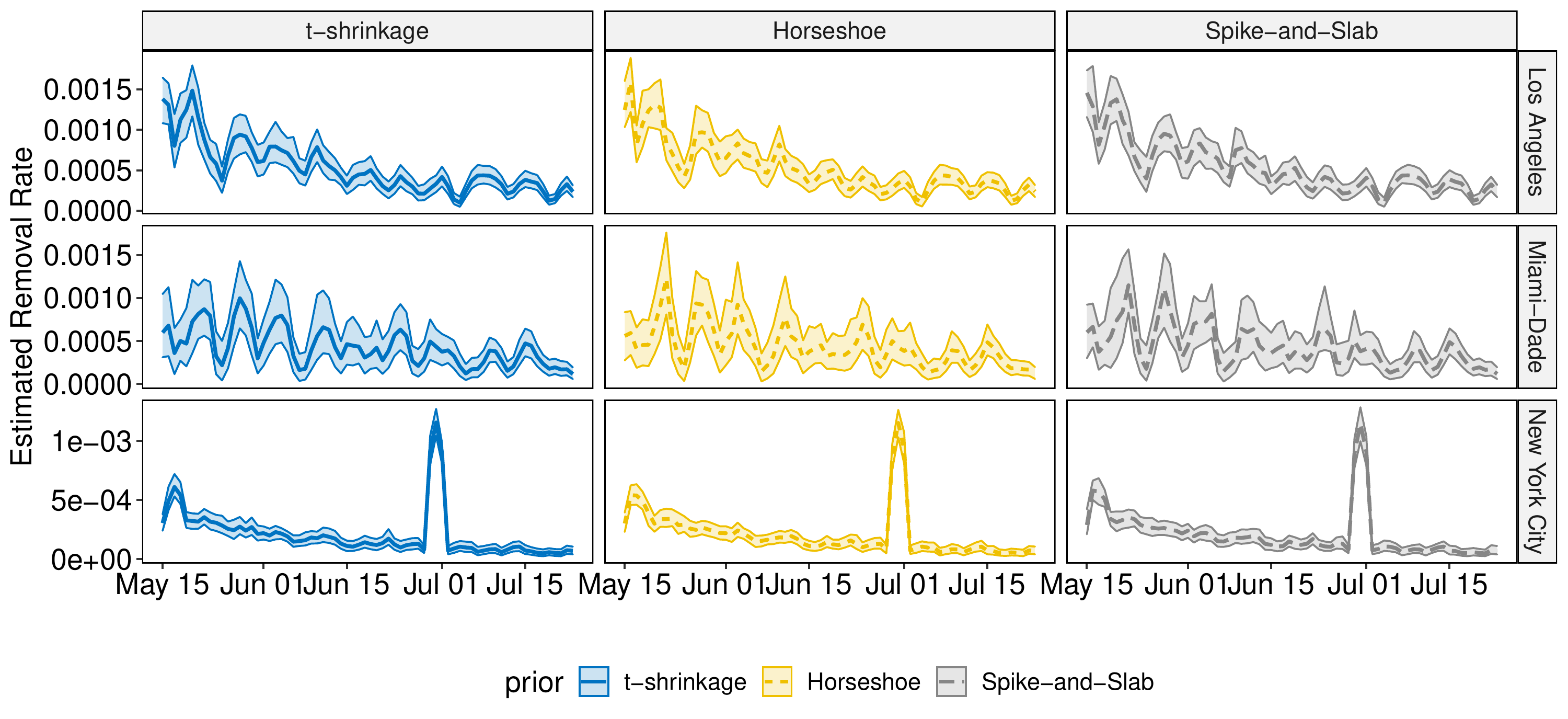}
	\caption{Plot for the estimated removal rate~$\gamma(t)$
	\blue{with 95\% HPD intervals}
	for the three selected counties over the studied
	time frame.}
	\label{fig:county_gamma}
\end{figure}

The proposed methodology is
applied on COVID-19 data for both state-level and county level introduced in
Section~\ref{sec:motivate_data}. Analysis for other states and counties can be
conducted in the same way, which is omitted in this paper.
\blue{Similar as the simulation studies, the chain length is set to 50,000
with thinning 10, and the first 3,000 samples after thinning are treated
as burn-in.}
%All
%other settings remain the same as simulation study. 
The estimated infectious rates and removal rates,
\blue{together with their 95\% highest posterior density (HPD) intervals}
are shown in Figures~\ref{fig:real_beta} to~\ref{fig:county_gamma}.

In both state-level and county-level, the three different priors yield
similar results. %We are interested in a fusion estimation for the trend. 
\blue{From Figures~\ref{fig:real_beta} and~\ref{fig:real_gamma}, }
we find the
infectious rate for NY is smaller than other states during this period.
\blue{Also, the relatively stable $\hat{\beta}(t)$ for NY after June~6th
indicates a potential cluster. }
FL witnesses a pump peak after mid June, which results from the aggressive
reopen in FL.
\blue{The clustered pattern in other states, however, is not as clear as
that in NY, which is partially due to the fact that testing and reporting are
conducted timely in NY as it is one of the initial hotspots in March and April
that experienced high growth of COVID-19 cases, while in the other states that
we considered, there is more delay in testing and reporting. As the number of
cumulative cases and daily new cases are large in CA, FL and NY, the HPD
band for $\hat{\beta}(t)$ is tight, while in SD and WY, where daily new cases
do not exceed, respectively, 100 and 50, the estimated HPD band is much wider.}
As for the removal rate, NY, FL and CA
have similar result. \blue{Despite the differences in numerical values, the
trends of $\hat{\gamma}(t)$ for NY and FL are similar, and display a weekly
seasonality, with the estimated removal rates being smaller than average on
weekends. This is due to the fact that reporting is less active during
weekends than during the week. Note that for SD, the estimated removal rate
clearly
shows a relatively stable pattern between June 20th and July 1st,
indicating the existence of a potential cluster. 
}
The nation-wide removal rate estimate is constant across the time
and has few peaks during this time period, \blue{as a few states release the
recovered cases in a cumulative manner on a certain day.}
%The trend of infectious and removal.
%rate for SD and WY are similar. 
%For county-level, the infectious
%of New York City looks flat.
\blue{On the county level, $\hat{\beta}(t)$ for New York City becomes stable
after June 15th. The infectious rate estimate for Miami-Dade remained
relatively low before June 22nd but started to increase, and remained
relatively high.
For Los Angeles, $\hat{\beta}(t)$ experiences fluctuations with weekly
seasonality, but the overall trend remains stable.}
%Miami-Dade county has higher infectious rate after
%mid-June. 
The \blue{estimated}
removal rates are very small in all three counties, \blue{and all have
wide HPD bands}, since the
removal group only contains deaths in county-level data. \blue{The sudden jump
in Figure~\ref{fig:county_gamma} for New York City corresponds to the release
of~633 death cases, which was due to data anomaly.}

\section{Conclusion}\label{sec:conclusion}

In this paper, we proposed the tf-SIR model to capture group structure for
infectious rate and removal rate of different time period by using Bayesian
shrinkage priors. To our best knowledge, this is the first attempt in
literature
to use Bayesian shrinkage to recover unknown grouping structure for SIR model.
Our simulation results indicate that the proposed method has reasonable
performance and ability to capture the group pattern of infectious rate and
removal rate. The analysis of COVID-19 data also brings in new understanding of
the infectious disease such as COVID-19. Also, our tf-SIR model can not only be
used to model and assess COVID-19 pandemic but also other epidemic.

\blue{One interesting consideration, as suggested by one anonymous reviewer,
is to use a supermartingale structure,
i.e., non-increasing function with respective to time,
to model the infectious rate~$\beta(t)$ because of the implementation of policy
intervention. While such assumption might not be met by the development of
COVID-19 in the United States, it is a quite reasonable assumption when
studying the course of development for COVID-19 in countries where quarantine
and stay-at-home policies are strictly enforced such as China,
Singapore, and Vietnam.
}

In addition, three topics beyond the scope of this paper are worth further
investigation. First, in our real data application, a moving average approach
is applied to deal with measurement errors for observed data. Proposing a
measurement error model with SIR is an interesting future work. Furthermore,
different states may have similar infection and removal pattern. Subgroup
detection for different states will help the government design its polices.
Finally, discovering theoretical guarantees such as posterior concentration
rates of proposed methods is also devoted to future research.

% \bibliographystyle{chicago}
% \bibliography{covid_19}

\end{document}